\documentclass[
  aps,
  prx,
  twocolumn,
  superscriptaddress,
  nofootinbib,
  nobibnotes,
  longbibliography,
  floatfix,
  10pt
]{revtex4-2}

\usepackage{graphicx}
\usepackage{amsmath}
\usepackage{amssymb}
\usepackage{bm}
\usepackage{booktabs}
\usepackage{makecell}
\usepackage{xcolor}
\usepackage{braket}
\usepackage[colorlinks=true,linkcolor=blue!60!black,citecolor=blue!60!black,
            urlcolor=blue!60!black]{hyperref}
\usepackage{cleveref}

\usepackage{changes}
\usepackage{orcidlink}
\definechangesauthor[name=Anthony, color=orange]{ag}

\definechangesauthor[name=Sanziana, color=blue]{ss}

\definechangesauthor[name=Ivano, color=brown]{ita}

\definechangesauthor[name=Alberto, color=green]{bai}

\DeclareUnicodeCharacter{2124}{\ensuremath{\mathbb{Z}}}

\newcommand{\Ztwo}{\mathbb{Z}_2}

\newcommand{\ZN}{\mathbb{Z}_N}
\newcommand{\nn}{\mathrm{nn}}

\newcommand{\Heff}{H_{\mathrm{eff}}}

\newcommand{\Papprox}{\mathcal{P}^{\mathrm{approx}}}
\newcommand{\dev}{\texttt{ibm\_phoenix}}

\usepackage{xcolor,colortbl}
\newcolumntype{a}{|>{\columncolor[gray]{0.8}}c}

\begin{document}


\title{Nonperturbative string dynamics in a $2+1$D lattice gauge theory on a superconducting quantum processor}

\author{S\^anziana Stelea}
\affiliation{IBM Research, 8803 R\"uschlikon, Switzerland}
\affiliation{Institute for Theoretical Physics, ETH Zurich, 8093 Zurich, Switzerland}

\author{Anthony Gandon}
\affiliation{IBM Research, 8803 R\"uschlikon, Switzerland}
\affiliation{Institute for Theoretical Physics, ETH Zurich, 8093 Zurich, Switzerland}

\author{Francesco Tacchino}
\affiliation{IBM Research, 8803 R\"uschlikon, Switzerland}

\author{Jad C. Halimeh${}^{\orcidlink{0000-0002-0659-7990}}$}
\affiliation{Department of Physics and Arnold Sommerfeld Center for Theoretical Physics (ASC),
Ludwig Maximilian University of Munich, 80333 Munich, Germany}
\affiliation{Max Planck Institute of Quantum Optics, 85748 Garching, Germany}
\affiliation{Munich Center for Quantum Science and Technology (MCQST), 80799 Munich, Germany}

\author{Ivano Tavernelli}
\affiliation{IBM Research, 8803 R\"uschlikon, Switzerland}

\date{\today}


\begin{abstract}
Electric flux strings underpin confinement in gauge theories. In two spatial dimensions, their evolution involves competing processes of charge motion, pair creation, and transverse deformation. Accessing these dynamics beyond perturbative limits challenges classical approaches, as entanglement growth restricts the times accessible to tensor-network simulations. Here we use quantum simulation to investigate a $2+1$D $\mathbb{Z}_2$ lattice gauge theory with dynamical matter and an explicit magnetic plaquette interaction, spanning strongly confined and nonperturbative deconfined regimes. Using 63 superconducting qubits, including 12 ancillas, we evolve initially localized strings with static or dynamical endpoint charges. Spatially resolved measurements reveal rapid charge and flux spreading in the deconfined regime, enhanced matter occupation in the string interior near the pair-creation resonance, and suppressed spreading at strong electric fields. String-path occupations further probe transverse rearrangements between configurations connected by plaquette operations. Retaining separate matter and gauge degrees of freedom makes local Gauss-law constraints directly accessible for error detection. Combining these constraints with ancilla checks and noise rescaling, we measure dynamics in circuits containing up to 15 second-order Trotter steps. Our results establish quantum simulation as a practical approach to exploring nonperturbative string dynamics in two spatial dimensions, with gauge redundancy serving as an experimental resource.

\end{abstract}

\maketitle

\section{Introduction}
\label{sec:intro}

Quantum simulation offers a route to probe real-time dynamics of gauge theories by encoding the
Hamiltonian directly into controllable quantum degrees of
freedom~\cite{Bauer2023,dimeglioQuantumComputingHighEnergy2024,halimeh2025quantumsimulationoutofequilibriumdynamics}.
Most experimental progress so far has been in one spatial dimension, where
models ranging from the Schwinger model to $\Ztwo$ and non-Abelian LGTs have
been realized on trapped
ions~\cite{martinezRealtimeDynamicsLattice2016,farrellPreparationsQuantumSimulations2023,deObservationStringbreakingDynamics2024},
ultracold
atoms~\cite{yangObservationGaugeInvariance2020,zhouThermalizationDynamicsGauge2022,Su2023}, and
superconducting
qubits~\cite{wangObservationEmergent22022,ciavarellaQuantumSimulationLattice2023,farrellQuantumSimulationsHadron2024,zemlevskiyScalableQuantumSimulations2025,schuhmacherObservationHadronScattering2025,alexandrouRealizingStringBreaking2025,mildenbergerConfinementZ2Lattice2025},
observing various phenomena ranging from string breaking and particle scattering to thermalization and its absence.

In two spatial dimensions ($2+1$D), the transverse spatial extent allows the flux tube to fluctuate and broaden,
enabling roughening and transverse string dynamics~\cite{luscherSymmetrybreakingAspectsRoughening1981,luscherHowThickAre1981,caselleEffectiveStringDescription2021,marcantonioRougheningDynamicsElectric2025,xuStringBreakingDynamics2025} and providing a more faithful picture of confinement than one-dimensional settings.
Observing these effects at experimentally accessible time scales requires an explicit magnetic plaquette term~\cite{tianRolePlaquetteTerm2025}, which deforms the string transversally by flipping the electric fluxes around a single plaquette.

Existing $2+1$D $\mathbb{Z}_2$ experiments fall into two classes. A first group realizes the gauge--matter model without a plaquette term~\cite{gyawaliObservationDisorderfreeLocalization2025,cobosRealTimeDynamics2+1D2025}. A second group includes the plaquette term but works in the \emph{gauge-fixed} (dual) formulation, in which the matter fields are eliminated by a local unitary and the model maps onto a toric code in external
fields~\cite{cochranVisualizingDynamicsCharges2025,xuObservationGlueballExcitations2026}. Beyond $\Ztwo$, a closely related confinement and string-breaking phenomenology has been observed for a dynamically generated $\mathbb{U}(1)$ gauge theory on a Rydberg-atom simulator~\cite{gonzalez-cuadraObservationStringBreaking2025}, as well as exactly on a trapped-ion platform~\cite{Joshi2026ObservationGenuine2D}, indicating that the string phenomenology targeted here is not specific to a single gauge group or hardware platform.
Gauge fixing is economical in qubits, but it discards precisely the structure exploited here: once the Gauss-law constraints have been solved, the local symmetry generators are no longer operators that can be measured on the device. Gauge-redundant encodings have been argued to outperform gauge-fixed ones below a finite physical error threshold~\cite{carenaQuantumErrorThresholds2024}.

Here, matter and gauge degrees of freedom are represented by distinct physical qubits, and the local Gauss operators $G_i$ survive as measurable stabilizers of the simulated state. The physical system we realize is a $5\times4$ matter-site lattice occupying $63$ qubits of the superconducting processor \dev, time-evolved for up to 15 Trotter steps. It supports 32 $\mathbb{Z}_2$ symmetries available for error detection: 20 originating from Gauss' law and 12 from the ancillas.

We focus on the time-dependent expectation values of diagonal observables that can be measured in the same basis as the system's symmetries. For observables that are additionally local, we show empirically that post-selection schemes based on marginal distributions, together with marginal error mitigation~\cite{gonzalesQuantumDistributionError2026,schuhmacherObservationHadronScattering2025} can replace global schemes. Over the time scales accessible experimentally, these incur a finite sampling overhead, which is offset by the fast $100~\textrm{kHz}$ sampling rate of the IBM Nighthawk R2 devices.

The paper is organized as follows. Section~\ref{sec:model} introduces the model
and the targeted string phenomenology, together with the digital implementation of its Trotter dynamics. Section~\ref{sec:results_quenches} presents the results of quantum-quench experiments across different regimes of the static phase diagram. Section~\ref{sec:discussion} discusses the outlook.

\begin{figure*}[ht!]
  \centering
  \includegraphics[width=1.0\linewidth]{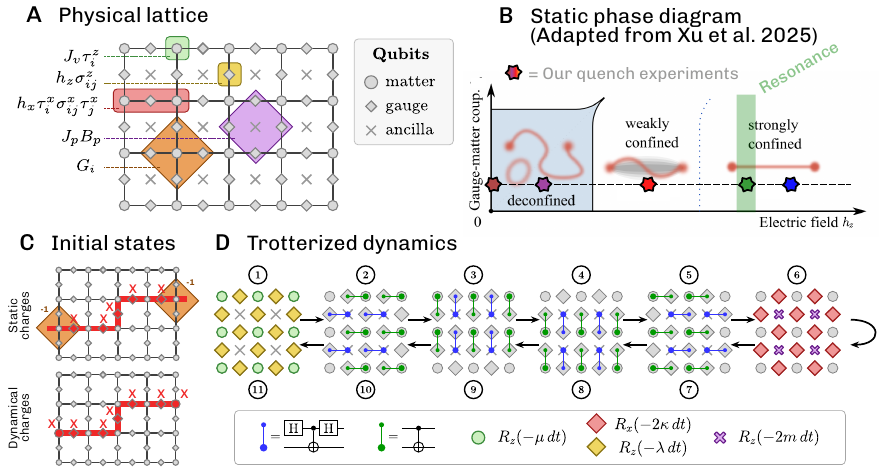}  
  \caption{\textbf{The $2+1$D $\Ztwo$ gauge theory on a square lattice with Ising matter}. (\textbf{A}) The physical lattice is composed of matter qubits on the vertices and gauge qubits on the links. Additional ancilla qubits are introduced to implement the four-body plaquette term (Sec.~\ref{sec:implementation}), resulting in a square qubit connectivity. The different highlighted regions mark the four Hamiltonian terms of
  Eq.~\eqref{eq:hamiltonian} and the Gauss-law constraint of
  Eq.~\eqref{eq:gauss}. (\textbf{B}) Quench experiments are realized in the different phases of the model. In the confined phase, a localized string typically survives over long time scales, while it quickly delocalizes in the deconfined regime.
  (\textbf{C}) Initial string state in the strongly confined regime $J_v/h_x=\infty$ with both static and dynamical endpoint charges. (\textbf{D}) Second-order Trotter dynamics as a depth-8 unitary evolution implementing the toric-code (TC) duality.}
  \label{fig:model}
\end{figure*}

\section{Model and quantum simulation setup}
\label{sec:model}

\textbf{\textit{Model---}} We consider the $\Ztwo$ lattice gauge theory with dynamical Ising matter, introduced by Fradkin and Shenker~\cite{fradkinPhaseDiagramsLattice1979}, on a two-dimensional square lattice. Matter degrees of freedom $\tau_i$ reside on the vertices $i$ and gauge degrees of freedom $\sigma_{ij}$ on the links $\langle ij\rangle$; both are two-level systems, so each maps onto a single qubit. The Hamiltonian reads 
\begin{equation}
  \label{eq:hamiltonian}
  H =
  - J_p \sum_p B_p
  - h_z \sum_{\langle ij \rangle} \sigma_{ij}^z
  - h_x \sum_{\langle ij \rangle} \tau_i^x \sigma_{ij}^x \tau_j^x
  - J_v \sum_i \tau_i^z .
\end{equation}
The first term is the plaquette operator $B_p=\prod_{\langle ij\rangle\in p}\sigma^x_{ij}$, measuring the magnetic $\Ztwo$ flux through an elementary plaquette, in analogy with $\oint \bm{A}\cdot d\bm{l}=\int \bm{B}\cdot d\bm{S}$ in electrodynamics. The second is the electric field energy, with the two eigenstates $\ket{0}$ and $\ket{1}$ indicating the absence and presence of $\mathbb{Z}_2$ electric flux.
The third term describes the minimal coupling between the gauge and matter degrees of freedom. It allows charges to move between neighboring sites and charge pairs to be created or annihilated, while simultaneously updating the connecting electric flux in accordance with Gauss's law, thereby rendering the matter dynamical.
The last term gives a mass to the matter fields.

The local $\Ztwo$ symmetry is generated by the Gauss operators
\begin{equation}
  \label{eq:gauss}
  G_i = \tau_i^z \prod_{j \in \nn(i)} \sigma_{ij}^z ,
  \qquad [H,G_i]=0 \ \ \forall i ,
\end{equation}
where $\nn(i)$ enumerates the links incident on site $i$. Equation \eqref{eq:gauss} is the lattice analogue of Gauss's law in electrodynamics: $\tau^z_i$ measures the $\Ztwo$ charge at site $i$, $\sigma^z_{ij}$ the electric field on the adjacent links,
and the constraint ties the product of the fields emanating from a site to the local charge. 
We consider sectors defined by \(G_i|\psi\rangle=q_i|\psi\rangle\), where \(q_i=\pm1\) specifies the static background-charge configuration. For strings with dynamical endpoints, \(q_i=+1\) at every site. For strings connecting two static sources, \(q_i=-1\) at the source positions and \(q_i=+1\) elsewhere. Since \([H,G_i]=0\), these sector labels are conserved under the ideal dynamics. The lattice, its interaction terms and Gauss operators are shown in Fig.~\ref{fig:model}(A).

The phase diagram of the model was studied analytically by Fradkin and Shenker~\cite{fradkinPhaseDiagramsLattice1979} and is summarized in Fig~\ref{fig:model}(B). It displays three regimes: a confined regime dominated by the electric field, a Higgs regime dominated by matter, and a deconfined regime dominated by the magnetic plaquette term. As detailed in Appendix~\ref{app:tc_duality}, The model maps onto the toric code in external fields under a local unitary transformation that integrates out the matter fields~\cite{wuPhaseDiagramToric2012,tupitsynTopologicalMulticriticalPoint2010}.

\paragraph*{Initial string states} In the strongly confined regime, $J_v/h_x=\infty$, the vacuum is the classical state $|\Omega\rangle = |00\dots\rangle$ with no matter and no electric field (assuming a periodic lattice). The lowest-lying excitations above this classical vacuum are string-like excitations along paths $\gamma$ connecting to matter endpoints $(i,j)$ and created by gauge-invariant operators of the form
\begin{equation}
  \label{eq:string_operator}
  \mathcal{O}_\gamma
  = \tau_i^x S_{\gamma} \tau_j^x ~~\text{with}~~ S_{\gamma}= \prod_{\langle kl \rangle \in \gamma} \sigma_{\langle kl \rangle}^x
\end{equation}
The energy of this electric flux string configuration is
\begin{equation}
  V(R) \approx 4\mu + 2\lambda R\,  ,
\end{equation}
which is linear in the charge separation $R$, so that in an infinite volume the charges cannot be separated at finite energy cost. With dynamical matter, $\kappa\neq0$, the string can instead break through the creation of a new charge pair, the lattice analogue of string breaking in QCD, where the color flux tube between heavy quark sources fragments through dynamical $q\bar q$ pair creation into color-neutral hadrons. Figure~\ref{fig:model}(C) illustrates the two string processes considered in this study.

\textbf{\textit{Quench dynamics setup---}} We study the dynamical aspects of the confined-deconfined phase transition in the Fradkin-Shenker model by repeating quench dynamics in different regimes of the phase diagram, as presented in Table~\ref{tab:parameter_regimes}.
\begin{table}[h]
\begin{tabular}{c||c|c|a|c|c|a}
        & $h_z$&$J_v$ &$h_z/J_v$ &$h_x$  & $J_p$ & $h_x/J_p$ \\ \hline
       deconfined & 0.00 & 1.00 & 0.00     & 0.25  & 1.0   & 0.25 \\ 
       deconfined & 0.25 & 1.00 & 0.25     & 0.25  & 1.0   & 0.25 \\ 
       weakly confined & 0.50 & 1.00 & 0.50     & 0.25  & 1.0   & 0.25 \\ 
       confined & 1.00 & 1.00 & 1.00     & 0.25  & 1.0   & 0.25  \\
       on resonance & 2.00 & 1.00 & 2.00     & 0.25  & 1.0   & 0.25 \\
       strongly confined & 4.00 & 1.00 & 4.00     & 0.25  & 1.0   & 0.25      
\end{tabular}
\caption{Parameter regimes}
\label{tab:parameter_regimes}
\end{table}

The quench dynamics consist of up to $15$ second-order Trotter steps with a time step of $dt=0.3$, for a total evolution time $t=4.5$ (in units of $J_v$), at which point the digital evolution remains an accurate approximation to the continuous-time dynamics. Each Trotter step comprises $8$ CNOT layers. For each evolution time, $10^6$ shots are collected and distributed over Pauli-twirling randomizations, which convert coherent gate errors into an effective stochastic Pauli channel without changing the logical circuit~\cite{wallmanNoiseTailoringScalable2016}.

\textbf{\textit{Post-selection and rescaling---}} Owing to hardware-level noise, the time-evolved quantum state accumulates errors in proportion to the total gate count and circuit depth. Viewed as distance-1 codes, $\Ztwo$ gauge theories possess error-detection capabilities that can be used to reduce the systematic bias of common observables, at the cost of an increased statistical variance. In addition, we apply a Clifford perturbation theory (CPT) rescaling of expectation values, based on noise factors learned by setting the non-Clifford angles to zero in the compiled circuits; this carries a sampling overhead proportional to the 1-norm of the estimated noise matrix, $\|A^{-1}\|_1$~\cite{schuhmacherObservationHadronScattering2025}. To limit the exponential sampling overhead of post-selection, the analysis is restricted to local, point-like observables and to approximations of string-like observables.

\textbf{\textit{Reference calculations. ---}}
\label{sec:peps_reference}
Out-of-equilibrium dynamics on a 2D square lattice are difficult to approximate classically, owing to the fast growth of the entanglement entropy~\cite{xuEntanglementPropertiesGauge2025}. For small global quenches, \emph{i.e.}, in the regime of large electric field, any deformation or splitting of the initial string is energetically penalized, and one can work at first order in perturbation theory in the reduced basis of minimal-length strings, for which exact diagonalization is available at the lattice sizes considered here. In the weakly confined and deconfined regimes, string elongation and splitting can no longer be neglected, and other classical approximations are required. Here we use a 2D PEPS tensor network~\cite{Verstraete2004RenormalizationAlgorithmsQuantumManya} to approximate the time-evolved wavefunction. Because of the high computational cost of contracting 2D tensor networks, we work at low bond dimension and report only the time steps for which convergence with respect to the bond dimension is satisfied, this is discussed further in~Appendix~\ref{app:simulability}.

\section{Results and string phenomenology}
\label{sec:results_quenches}
\subsection{Matter sector}

\begin{figure*}[ht!]
    \centering
    \includegraphics[width=\textwidth]{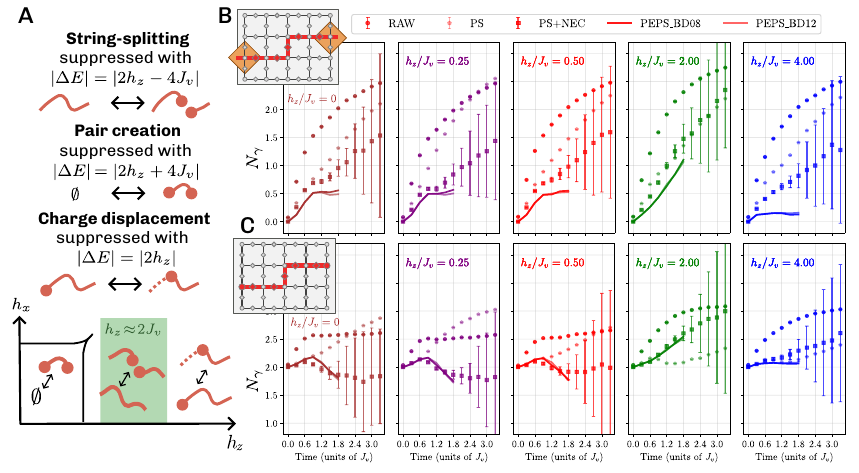}
    \caption{\textbf{Matter occupation numbers across the initial string}. (\textbf{A}) Channels acting on the matter occupation numbers. At lowest order in perturbation theory, these channels are weighted by the inverse energy difference of the initial and final configuration. (\textbf{B}) Time dynamics of the total matter occupation along a gauge-violating initial string with endpoint static charges. The raw and mitigated hardware data are shown in comparison with the PEPS reference result, when converged and available. (\textbf{C}) Same experiment with dynamical endpoint charges. In this case, the dynamics along the string in the deconfined regime are dominated by the displacement of the initial charges.}
    \label{fig:static_string}
\end{figure*}

\textbf{\textit{String with static endpoints.---}} Starting from the product state \(|\Omega\rangle=|00\ldots0\rangle\), we prepare an electric string along the path \(\gamma\) shown in Fig.~\ref{fig:static_string}(B) by applying the gauge-variant $S_\gamma$ operator of Eq.~\ref{eq:string_operator}.
The resulting state contains electric flux along \(\gamma\) and no dynamical matter excitations. It belongs to the sector with \(q_i=-1\) at the two endpoints and \(q_i=+1\) elsewhere, corresponding to two static background charges connected by the string. The source positions remain fixed throughout the evolution. This preparation removes the charge-displacement channel present when the endpoints are dynamical, while still allowing subsequent matter-pair creation, string breaking, and motion of the created charges.

The gauge–matter coupling can create dynamical charge pairs both on occupied string links and on initially empty links. The corresponding energy changes follow from the diagonal part of Eq.~\eqref{eq:hamiltonian}. Acting with \(\tau_i^x\sigma_{ij}^x\tau_j^x\) on an occupied link whose two adjacent matter sites are empty creates two matter excitations, costing \(4J_v\), while removing one unit \(2h_z\) of electric flux. The net energy change is therefore \(\Delta E=4J_v-2h_z\). This process breaks the string and has a bare resonance at \(h_z=2J_v\), irrespective of whether its endpoints are static sources or dynamical charges. By contrast, creating a pair on an empty link also creates electric flux and costs \(\Delta E=4J_v+2h_z\), which becomes energetically more expensive deeper in the confined regime.

Figure~\ref{fig:static_string}(B) shows the resulting total matter occupation 
$N_{\gamma} = \sum_{i\in\gamma} n_i$ as a function of time for the six ratios $h_z/J_v$ of Table~\ref{tab:parameter_regimes}, together with the PEPS reference where the bond-dimension scan has converged. In every regime the occupation grows monotonically from zero, confirming that pair creation along the string remains active throughout, consistent with the detuning argument above: the deconfined points ($h_z/J_v=0$ and $0.25$) show the fastest matter buildup, while the strongly confined point ($h_z/J_v=4$) is markedly slower. The PEPS curves are regarded as reliable only over the time window in which convergence has been checked (Sec.~\ref{sec:peps_reference}). In the regime where it has converged, the PEPS prediction tracks the raw and mitigated hardware data at early times. At later times, the hardware data effectively converges to an almost fully mixed state and mitigation leads to large error bars. Overall, the comparison shows that the gauge-violating string geometry isolates the pair-creation channel cleanly, giving a monotonic, regime-dependent growth rate that is complementary to the non-monotonic, resonance-driven dynamics of the dynamical-endpoint string discussed below.

\begin{figure*}[t]
    \centering
    \includegraphics[width=\textwidth]{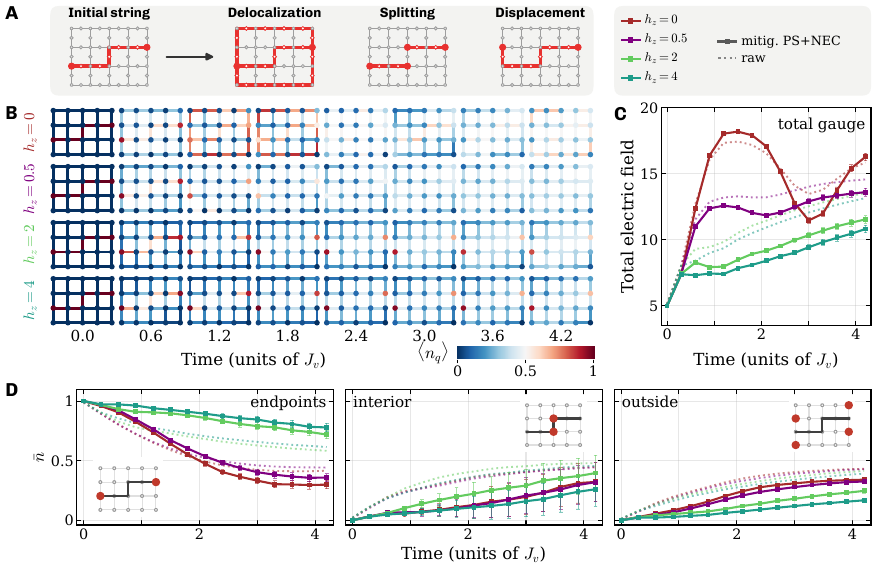}
    \caption{\textbf{Dynamics of a string system with dynamical endpoints}. (\textbf{A}) The initial string on the $5\times4$ lattice and the main phenomenological effects occurring across the phase diagram. In delocalization, new flux is created all across the lattice; in splitting, a charge pair is created on the string, breaking it; in displacement, an endpoint charge moves to a neighboring site. (\textbf{B}) Mitigated onsite occupation of all matter sites and gauge links across the four parameter regimes, for the subset of time steps $[0, 0.6, 1.2, 1.8, 2.4, 3.0, 3.6, 4.2]$. (\textbf{C}) Total gauge occupation of the lattice. (\textbf{D}) Mean matter occupation at the two initial endpoints, at the two central sites of the initial string (interior), and at the four sites adjacent to the endpoints but off the string (outside).}
    \label{fig:density_plots}
\end{figure*}

\textbf{\textit{String with dynamical endpoints.---}} The same experiment with a gauge-invariant path $\gamma$ with dynamical charges at its endpoints is reported in Fig.~\ref{fig:model}(C), prepared by acting with the operator $\mathcal{O}_\gamma$ of Eq.~\eqref{eq:string_operator} on the vacuum state of a $5\times4$ lattice of matter sites. In the deconfined regime, the behavior differs markedly: the initially inserted charges eventually disperse throughout the lattice, leading to a drop in the string occupation $N_\gamma$.

Characterizing the string dynamics when the endpoints are not fixed, as in the phenomenology outlined in Fig.~\ref{fig:density_plots}(A), requires observables defined on the full $5\times4$ square lattice. Panel (B) displays, for a subset of four values of the electric field $h_z\in [0, 0.5, 2, 4]$, the measured onsite occupation of every matter and gauge qubit of the lattice at each of these times, with color encoding the local occupation number from $0$ (blue) to $1$ (red); the spreading of this occupation away from the initial string directly visualizes the widening and delocalization of the initial flux tube.

Deep in the deconfined regime ($h_z=0$), the string does not remain a well-defined object: additional electric flux is created, its occupation spreads rapidly over the lattice and is subject to strong oscillations until late times (Fig. \ref{fig:density_plots}(C)). In the deep confined regime, $h_z=4$, the string instead remains strongly localized, and its widening is limited to a very small extent, staying visually close to its initial shape throughout the evolution shown in panel B. The chosen initial string shape is degenerate, in the strongly confined regime, with 5 strings of equal length, which gives the string a small transversal extent at equilibrium, even in the confined phase, and hence small oscillations out of equilibrium.

\begin{figure*}[ht]
    \centering
    \includegraphics[width=\linewidth]{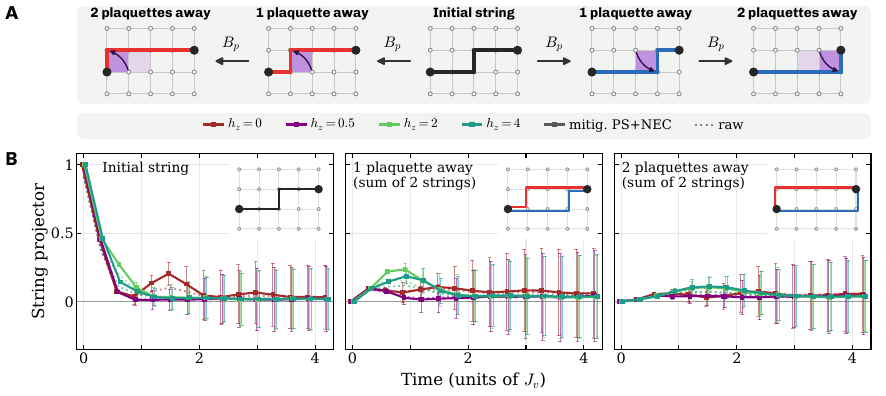}
    \caption{\textbf{Partial string-like projectors}, grouped according to perturbation theory. (\textbf{A}) Overview of minimal-length string oscillations driven by the plaquette term. (\textbf{B}) Measured string projector $\langle\prod_{k\in\gamma} n_k\rangle$ on the links of each string $\gamma$, for the initial string (left), summed over the two strings 1 plaquette (center) and 2 plaquettes (right) away, for the four values of $h_z$.}
    \label{fig:projectors}
\end{figure*}

At the resonance $h_z=2J_v$, the principal channel for matter creation is string splitting along the initial path $\gamma$, due to the energy difference scaling as $|1/\Delta E_{\rm splitting}| \propto 1/|h_z-2J_v|$. This process is suppressed away from resonance. Instead, displacement of the initial charges becomes dominant at $h_z=0$, with a scaling $|1/\Delta E_{\rm disp}| \propto 1/|h_z|$. To quantify this effect, we measure the matter occupation of the string's initial endpoints, of its two most central sites, as well as of the sites outside the string that neighbor its endpoints.
\begin{gather}
    \bar n_{\rm end} = \frac{1}{2}\left(n_i + n_j\right), \\
    \bar n_{\rm int} = \frac{1}{2}\sum_{k\in \mathcal{C}} n_k, \qquad
    \bar n_{\rm out} = \frac{1}{4}\sum_{k\in \mathcal{N}} n_k,
\end{gather}
where $i=(0,1)$ and $j=(4,2)$ are the initial endpoints of the string,
$\mathcal{C}=\{(2,1),(2,2)\}$ are its two central sites, and
$\mathcal{N}=\{(0,0),(0,2),(4,1),(4,3)\}$ are the sites outside the string
that neighbor its endpoints.
The raw and mitigated hardware data for both observables are shown in Fig.~\ref{fig:density_plots}(D). The endpoint occupation $\bar n_{\rm end}$ decays monotonically from its initial value of $1$ in all four regimes, fastest at $h_z=0$ and $h_z=0.5$ and markedly slower at $h_z=2$ and $h_z=4$, while the interior occupation $\bar n_{\rm int}$ grows from $0$, most rapidly at the resonance $h_z=2$. This is consistent with the perturbation-theory argument above: the resonant case is characterized by fast matter creation in the interior of the string, \emph{i.e.}, proportional to its length, while the confined phase away from resonance is dominated by point-like effects at the endpoints. In the deconfined regime, the decay of the endpoint occupation is accompanied by growth of both $\bar n_{\rm int}$ and $\bar n_{\rm out}$, which reach comparable values at late times, indicating that the dominant process is displacement of the initial charges, which spread over the lattice with no preference for the sites of the initial string. The two channels for matter creation, charge displacement and string splitting, thus leave distinct and separately measurable signatures in the local matter occupation.


\subsection{Gauge sector}
\textbf{\textit{Delocalization and string widening.---}}
During the quench dynamics, initial string configurations oscillate owing to the magnetic plaquette term.
The amplitude of each string configuration is typically quantified with Loschmidt echoes, which are global projection operators. To keep the observables compatible with the local post-selection and rescaling scheme, we instead consider partial string-like projectors
\begin{equation}
    \label{eq:string_projector}
    P_\gamma = \prod_{ij\in \gamma} \frac{1 - \sigma_{ij}^z}{2}
\end{equation}
for various paths $\gamma$ connecting the initial charge locations. In the strongly confined regime, the vanishing of the initial string probability observed in Fig.~\ref{fig:density_plots}(B) is followed by an increase in the amplitudes of the other equal-length configurations, as shown in Fig.~\ref{fig:projectors}(B). The total probability of equal-length strings instead is expected to reach a steady value that is independent of the particular choice of string configuration.

\section{Discussion and outlook}
\label{sec:discussion}

We have implemented the real-time dynamics of a $2+1$D $\Ztwo$ lattice gauge theory with dynamical Ising matter and an explicit plaquette term on $63$ qubits of a superconducting processor, retaining the gauge-redundant encoding so that the local Gauss-law generators remain measurable. The matter creation channels of the model were probed using the total matter occupation along the initial string, revealing distinct string dynamics in the deconfined, weakly confined and strongly confined regimes.
In the confined regime, matter creation is mostly present at resonances and concentrated along the initial string. In the deconfined regime, we observe both matter creation and displacement of the initial charges across the full extent of the lattice, largely independently of the electric field strength.
 The gauge content of the theory was probed separately. In the confined regime, the electric flux was observed to remain qualitatively localized close to the subset of minimal-length strings connecting the initial charges, with only a small transverse extent at early times, whereas deep in the deconfined regime it was observed to spread rapidly across the lattice and to oscillate strongly. The oscillations of the confined string toward the other minimal-length configurations, driven by the plaquette term, were further probed quantitatively through partial string-like projectors.

Methodologically, local post-selection combined with a rescaling of hardware expectation values is found to accurately reproduce PEPS results converged up to 7 Trotter steps. At longer times, convergence of the PEPS, and of the resulting expectation values, becomes slow for bond dimensions larger than 8; this finite-bond-dimension error is especially large in the deconfined regime, \emph{i.e.}, for large quantum quenches, owing to the rapid growth of entanglement~\cite{xuEntanglementPropertiesGauge2025}. The rescaling approach used here instead relies on Clifford perturbation theory, and is better suited to quantum circuits with small electric field strength and correspondingly small non-Clifford Trotter angles.

Gauge redundancy---usually treated as overhead to be removed by gauge fixing---supplies local, measurement-basis-diagonal stabilizers evaluated for free from the samples already collected. Notably, the post-selection rate is found to be largely independent of the quench parameters at the present system size. Mitigation and detection are complementary: mitigation can in principle be unbiased, but its overhead grows rapidly with depth, whereas detection is cheaper and non-exponential for local constraints~\cite{yuanZenoEnhancedProbabilisticError2026}. The circuit structure enables further extensions: mid-circuit syndrome extraction, measurement-based gauge protection via a predicted quantum Zeno transition~\cite{wautersSymmetryprotectionZenoPhase2025}, and measurement-free alternatives (energy penalties~\cite{halimehGaugeSymmetryProtectionUsing2021,halimehReliabilityLatticeGauge2020,halimehStabilizingGaugeTheories2022,lammSuppressingCoherentGauge2020,nguyenDigitalQuantumSimulation2022,cobosRealTimeDynamics2+1D2025}) are all available at no additional syndrome-extraction overhead, since the Gauss and ancilla stabilizers are already encoded into single qubits at various points of the Trotter step (Appendix~\ref{app:propagation}). Pairwise Gauss-stabilizer violations can also be attributed to a single gauge-qubit bit flip with high confidence, so correcting rather than discarding such samples would extend the accessible evolution time. This remains relevant in the era of partial fault tolerance, and points toward larger gauge groups $\ZN$ and richer matter content, in particular fermions.

\begin{acknowledgments}
This work was carried out at IBM Research Europe -- Zurich. We thank the IBM Quantum team for access to and support with the \dev~processor. Zurich. J.C.H.~acknowledges funding by the Max Planck Society, the Deutsche Forschungsgemeinschaft (DFG, German Research Foundation) under Germany's Excellence Strategy – EXC-2111 – 390814868, and the European Research Council (ERC) under the European Union's Horizon Europe research and innovation program (Grant Agreement No.~101165667)—ERC Starting Grant QuSiGauge. J.C.H.~acknowledges the hospitality of the Aspen Center for Physics, which is supported by National Science Foundation grant PHY-2210452, while this paper was being completed.
This work is part of the Quantum Computing for High-Energy Physics (QC4HEP) working group. The authors acknowledge the usage of ChatGPT and Claude to improve the readability of the manuscript.
\end{acknowledgments}

%

\newpage
\onecolumngrid
\appendix
\setcounter{figure}{0}
\renewcommand{\thefigure}{A\arabic{figure}}
\renewcommand{\theHfigure}{A\arabic{figure}}

\section{Quantum hardware}

\begin{figure}[h!]
  \centering
  \includegraphics[width=0.8\linewidth]{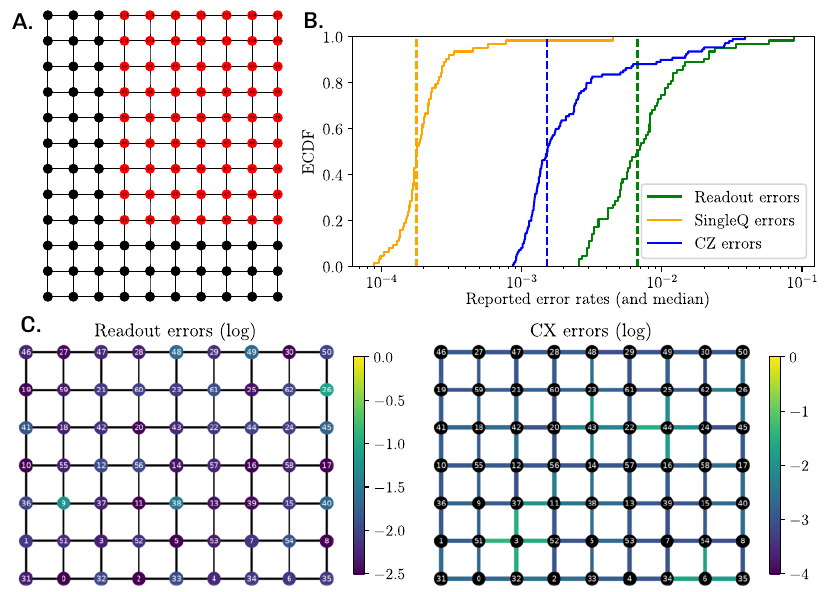}
  \caption{\textbf{Calibration property of the quantum processor \dev}. (\textbf{A}) Embedding of the 5x4 physical system in the square lattice qubit hardware. (\textbf{B}) Cumulative distribution of error rates for single qubit gates, entangling CZ gates and measurement errors. (\textbf{C}) Visualization of error rates per qubits and per CZ gate.}
  \label{fig:calibration_data}
\end{figure}

For the present study we used a 9$\times$7 sublattice of the full 10$\times$12 \dev~processor. The selected qubits as well as the calibration data for the gate errors are all reported in Fig~\ref{fig:calibration_data}. For the experiments ran in the main text, the median error rates are: $6.7\times 10^{-3}$ for the readout errors, $1.8\times 10^{-3}$ for the single qubit gate errors, and $1.5\times 10^{-3}$ for the CZ entangling gate errors.

\section{Digital implementation and its stabilizer structure}
\label{sec:implementation}
In this section we discuss in more details the digital implementation of the Trotter dynamics in the Fradkin and Shenker model.

\subsection{Trotter circuit}

We split the Hamiltonian into two non-commuting parts, $H=H_A+H_B$, with
\begin{equation}
  H_A = -h_z \sum_{\langle ij \rangle} \sigma_{ij}^z - J_v \sum_i \tau_i^z ,
\end{equation}
\begin{equation}
  H_B = -J_p \sum_p \prod_{\langle ij \rangle \in p} \sigma_{ij}^x
        - h_x \sum_{\langle ij \rangle} \tau_i^x \sigma_{ij}^x \tau_j^x ,
\end{equation}
and approximate the evolution operator by a second-order Suzuki--Trotter product
\begin{equation}
  \label{eq:trotter}
  e^{-iHt} \approx \big( U_A(dt/2) U_B(dt) U_A(dt/2 \big)^N ,
  \quad \Delta t = t/N ,
\end{equation}
with $U_A(dt/2)=e^{-iH_A dt/2}$ and $U_B(dt=e^{-iH_B dt}$.

$U_A$ consists of single-qubit $Z$ rotations on all gauge and matter qubits and
executes in parallel across the lattice. $U_B$ contains the four-body plaquette
and three-body gauge--matter interactions. Following the ancilla-mediated
construction of Ref.~\cite{zoharDigitalLatticeGauge2017}, the plaquette term is implemented with one ancilla qubit at the center of each plaquette: each of the four surrounding gauge qubits is entangled with the ancilla by an $H$--CNOT--$H$ sequence (gauge qubit as control, ancilla as target), a single-qubit $R_z$ on the ancilla generates the plaquette phase, and the sequence is reversed to disentangle the ancilla. The hopping term is implemented by CNOTs from the matter qubits onto the gauge qubit, a single-qubit $R_x$ on the gauge qubit, and the reversed CNOT sequence. The two terms are implemented in parallel. The full step is shown in Fig.~\ref{fig:model}(A).

Apart from the single-qubit rotations proportional to the Trotter time step $dt$ in Fig~\ref{fig:model}, every operation in the circuit belongs to
the Clifford group. Removing the rotations leaves a purely Clifford circuit which, in our construction, reduces trivially to the identity, since all entangling gates appear in forward/reverse pairs. This provides a classically trivial noise-estimation circuit, exploited in Sec.~\ref{sec:mitigation}.

\subsection{Operator propagation and stabilizers}

To see how the circuit encodes the local interactions, it is useful to track Pauli operators through the entangling Clifford layers in the Heisenberg
picture. For a CNOT with control $c$ and target $t$,
\begin{equation}
  X_c \to X_c X_t, \quad Z_c \to Z_c, \quad
  X_t \to X_t, \quad Z_t \to Z_c Z_t ,
\end{equation}
while for the dressed gate
$U=(H_c\otimes I_t)\,\mathrm{CNOT}\,(H_c\otimes I_t)$,
\begin{equation}
  X_c \to X_c, \quad Z_c \to Z_c X_t, \quad
  X_t \to X_t, \quad Z_t \to X_c Z_t .
\end{equation}
Applying these rules for the Toric code unitary disentangler of Fig~\ref{fig:model}, we find that all gauge stabilizer take a single qubit form at the middle of each Trotter step circuit.

Additionally, the ancillas are prepared as $|0\rangle$, \emph{i.e.} as trival stabilizer states and in the absence of errors they return to
$|0\rangle$ after each step. The total state then has for stabilizer group
\begin{equation}
  \label{eq:stabilized_subspace}
  \mathcal{S} =
  \big\{ |\psi\rangle \;\big|\;
    G_i|\psi\rangle=|\psi\rangle,\;
    Z_a|\psi\rangle=|\psi\rangle,\;\forall i,a \big\} .
\end{equation}
At the midpoint of the step the same stabilizer group is represented differently: the Gauss stabilizers become single-qubit $Z$ operators on the matter qubits, and the ancilla stabilizers become five-body operators consisting of $Z_a$ on the ancilla and $X$ on the
four surrounding gauge qubits, see Fig.~\ref{fig:model}(D). Both representations
are local, and both are available at natural points of the circuit without any
additional gates.

Crucially, the end-of-step stabilizers are diagonal in the computational basis.
Since all our measurements are performed in that basis, their eigenvalues can be
computed directly from each measured bitstring. The syndrome is therefore free:
it requires no syndrome-extraction circuit, no extra qubits beyond the plaquette
ancillas already needed for the dynamics, and no mid-circuit measurement.

\subsection{Toric code duality}
\label{app:tc_duality}

The Hamiltonian in Eq.\ref{eq:hamiltonian} is unitarily related to Kitaev's toric code in an external field~\cite{kitaevFaulttolerantQuantumComputation2003},
$H_{\rm TC}=-J_v\sum_sA_s-J_p\sum_pB_p - h_x \sigma^x_{ij}  - h_z \sigma^z_{ij}$ with
$A_s=\prod_{\langle ij\rangle\in\mathrm{star}(s)}\sigma^z_{ij}$ and
$B_p=\prod_{\langle ij\rangle\in p}\sigma^x_{ij}$. With no external fields, the generators commute and generate
the stabilizer group whose $+1$ eigenspace is the code space, coinciding with
the ground-state manifold. Local Pauli errors create pairs of stabilizer
defects: an $X$ error anticommutes with neighboring vertex stabilizers and
creates electric excitations, a $Z$ error creates magnetic ones through violated
plaquettes. In the anyonic language these defects are quasiparticles, and the
propagation of anyons maps onto the propagation of error strings. Since
stabilizer measurements reveal only the endpoints of an error chain, decoding
amounts to reconstructing the most likely chain from the syndrome---so the
dynamics of excitations in $\Ztwo$ gauge theories is directly relevant to
decoder design. The \emph{surface code} is the planar realization with open
boundaries, in which vertex and plaquette operators are truncated at the
boundary~\cite{fowlerSurfaceCodesPractical2012}; its locality and high threshold
make it a leading candidate for fault-tolerant
computation~\cite{stephensFaulttolerantThresholdsQuantum2014,acharyaQuantumErrorCorrection2025,satzingerRealizingTopologicallyOrdered2021}.

In our un-gauge-fixed model the analogous role is played by the Gauss operators
$G_i$, Eq.~\ref{eq:gauss}. A bit flip on a matter qubit
$\tau_i$ creates a single stabilizer defect at $i$; a bit flip on a gauge qubit
$\sigma_{ij}$ creates two defects at the neighboring sites $i$ and $j$. Because
the symmetry is discrete, $G_i^2=\openone$ and the Hilbert space splits into
binary sectors, so gauge-violating errors appear as binary syndrome defects
closely resembling those of stabilizer codes. This differs qualitatively from
LGTs with continuous gauge symmetry: U$(1)$ quantum link
models~\cite{chandrasekharanQuantumLinkModels1997} are characterized by additive
electric-flux divergence constraints, so gauge-violating errors change local
charge sectors rather than flipping binary labels, and---although spin-$\tfrac12$
quantum link models remain qubit-encodable---the resulting syndrome structure is
less directly compatible with standard stabilizer decoding. Fault-tolerant
protocols exploiting the native stabilizer structure of LGTs are an active
direction~\cite{strykerOraclesGausssLaw2019,rajputQuantumErrorCorrection2023,spagnoliFaulttolerantSimulationLattice2026,spagnoliQuditStabiliserCodes2026,gustafsonRobustnessGaugeDigitization2023,yaoQuantumErrorCorrection2025,patoTradeoffsGausssLaw2026,carenaQuantumErrorThresholds2024}.

\section{Local error detection and local error mitigation}
\label{sec:mitigation}

Noise on current hardware drives the state out of the stabilized subspace
$\mathcal{S}$ of Eq.~\eqref{eq:stabilized_subspace}. Full quantum error
correction is out of reach, but the stabilizer structure can be exploited
through post-selection, and the Clifford structure of the circuit through
distribution-based mitigation. We describe both, and then their combination.

\subsection{Local stabilizer post-selection}
\label{sec:postselection}

The stabilizers are sensitive to isolated single-qubit errors; simultaneous
errors on neighboring qubits can leave the local eigenvalues unchanged and
remain undetected. As the circuit deepens, errors accumulate and it becomes
increasingly likely that several nearby qubits are affected, so the efficacy of
stabilizer-based post-selection degrades with evolution time.

Given the circuit structure, the Gauss stabilizers are primarily sensitive to
bit-flip errors, whereas the ancilla stabilizers provide sensitivity to
phase-flip errors (Appendix~\ref{app:propagation}). Because all measurements are
in the computational basis, phase flips are not directly visible in the final
outcomes, so Gauss-law post-selection is expected to contribute more strongly to
the recovery of the physical signal. We nonetheless use both.

\paragraph*{Global post-selection.} Retaining only samples that satisfy every
constraint is immediately impractical at our system size. For the $5\times 4$
lattice this means $20$ Gauss constraints and $12$ ancilla constraints.

\paragraph*{Local post-selection.} The locality of the stabilizers circumvents
this exponential overhead. Rather than enforcing all constraints simultaneously
across the lattice, we post-select on reduced local subsystems. This is
physically motivated: the observables are local, and error propagation during
finite-time evolution is approximately confined within an effective light cone
\cite{liebFiniteGroupVelocity1972}, so gauge violations far from the support of
an observable have little influence on its expectation value.

\begin{figure*}[ht!]
  \centering
  \includegraphics[width=\linewidth]{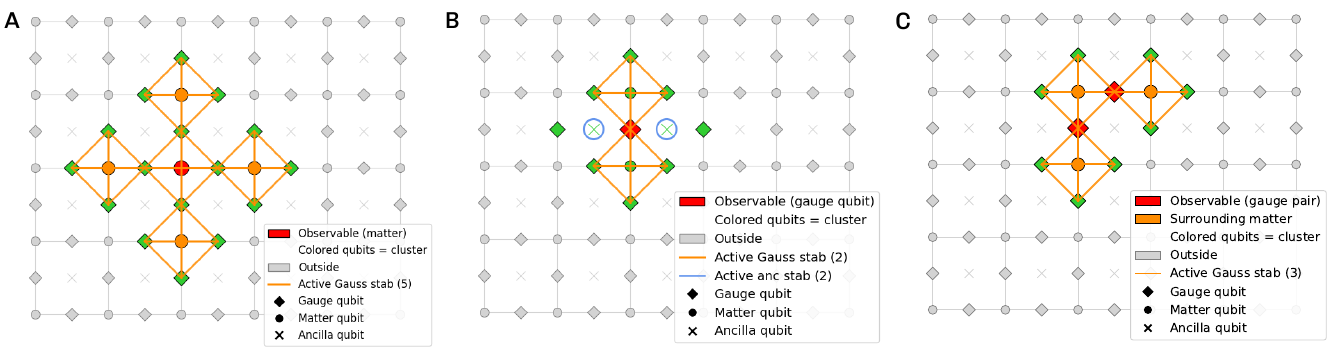}
  \caption{\textbf{Local clusters used for post-selection and error mitigation}. (\textbf{A}) The local cluster used for the matter occupations $t_{i}$, with $5$ included Gauss conditions. (\textbf{B}) Cluster for a single link occupation $\sigma_{ij}$, containing $2$ Gauss and $2$ ancilla stabilizers. (\textbf{C}) Cluster for a pair projector $\langle P_iP_j\rangle$ of neighboring links, containing $3$ Gauss stabilizers.
  }
  \label{fig:postselection}
\end{figure*}

For each observable we define a cluster containing the observable and its
neighborhood, chosen to maximize the number of locally accessible stabilizer
constraints while keeping the subsystem small enough to retain an appreciable
yield.  
For a single link occupation
$\langle \sigma_{ij}\rangle$ we use Fig.~\ref{fig:postselection}(B), with $2$ Gauss and
$2$ ancilla stabilizers. For the matter occupations we use a cluster including $5$ neighboring Gauss stabilizers of the local matter neighborhood.



\subsection{Combining post-selection with error-mitigation}
\label{sec:combination}

Post-selection discards corrupted samples. To additionally correct the retained distribution we use distribution error mitigation (DEM)~\cite{gonzalesQuantumDistributionError2026} in its marginal variant~\cite{schuhmacherObservationHadronScattering2025}.

Because both techniques are local and act on the same marginal distributions, they can be composed. There is also a practical reason to combine them: error mitigation generally carries a sampling overhead that grows exponentially with depth as noise accumulates~\cite{yuanZenoEnhancedProbabilisticError2026}, whereas post-selection at these system sizes has a comparatively mild overhead (Fig.~\ref{fig:postselection}(g)). 
Applying post-selection first makes the reconstruction act within the stabilized subspace, so the effective noise model seen by DEM is restricted to the Pauli processes that preserve the stabilizers; the post-selection then effectively tailors the channel entering the reconstruction. This conditioning does not break the block-circulant structure of $A$, so the procedure remains well defined either way.

\subsection{Statistical error bound for the distribution reconstruction}
\label{app:bound}

For a diagonal observable $\langle O\rangle=\sum_x O(x)p(x)$ estimated from $N$
shots, standard Monte Carlo gives
$\mathrm{Var}(\widehat O)=\frac1N(\mathbb E_p[O^2]-\mathbb E_p[O]^2)$. In Clifford-perturbation based rescaling methods~\cite{schuhmacherObservationHadronScattering2025,farrellQuantumSimulationsHadron2024}, the
reconstructed distribution is
$\widetilde p(x)=\sum_y(A^{-1})_{xy}p_{\mathrm{noisy}}(y)$, so
\begin{equation}
  \langle O\rangle_{\mathrm{DEC}}
  = \sum_y\Big[\sum_x O(x)(A^{-1})_{xy}\Big] p_{\mathrm{noisy}}(y) .
\end{equation}
Defining the reweighted observable
$O_{\mathrm{DEC}}(y):=\sum_x O(x)(A^{-1})_{xy}$, we have
$\langle O\rangle_{\mathrm{DEC}}
 =\mathbb E_{p_{\mathrm{noisy}}}[O_{\mathrm{DEC}}]$ and
\begin{align}
  \mathrm{Var}(\widehat O_{\mathrm{DEC}})
  &= \frac1N\Big(\mathbb E_{p_{\mathrm{noisy}}}[O_{\mathrm{DEC}}^2]
    - \mathbb E_{p_{\mathrm{noisy}}}[O_{\mathrm{DEC}}]^2\Big) \nonumber\\
  &\leq \frac1N \mathbb E_{p_{\mathrm{noisy}}}[O_{\mathrm{DEC}}^2] .
\end{align}
If $|O(x)|\leq O_{\max}$ then
$|O_{\mathrm{DEC}}(y)| \leq O_{\max}\sum_x|(A^{-1})_{xy}|$, so with the matrix
$1$-norm $\|A^{-1}\|_1=\max_y\sum_x|(A^{-1})_{xy}|$ the standard deviation obeys
\begin{equation}
  \sigma_{\mathrm{DEC}} \leq \frac{O_{\max}\|A^{-1}\|_1}{\sqrt N} .
\end{equation}
All observables in this work have $O_{\max}=1$.


\subsection{Error mitigation benchmark on a \texorpdfstring{$3\times3$}{3x3} lattice}
\begin{figure*}[h!]
    \centering
    \includegraphics[width=\textwidth]{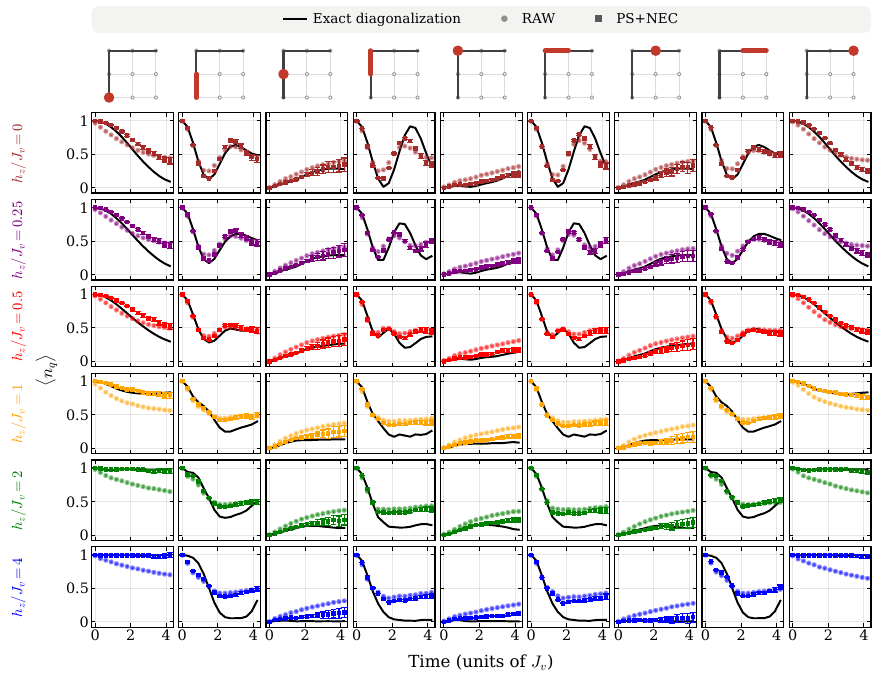}
    \caption{\textbf{Benchmark of the error mitigation against exact diagonalization} on the $3\times3$ lattice with an L-string instead of an horizontal string. Occupation $\langle n_q\rangle$ of each matter site and gauge link along the initial string (columns, highlighted in the sketches), for the six values of $h_z/J_v$ (rows). Black lines show exact continuous-time evolution, circles the raw hardware data, and squares the data after local post-selection and rescaling, using the gauge and matter clusters of Sec.~\ref{sec:postselection}.}
    \label{fig:ed_benchmark_3x3}
\end{figure*}

To benchmark the error mitigation against an exact reference, we run hardware experiments on a $3\times3$ lattice of matter sites for the six ratios $h_z/J_v$ of Table~\ref{tab:parameter_regimes}, starting from an L-shaped string connecting opposite corners of the lattice. The raw and mitigated occupations along the initial string are shown in Fig.~\ref{fig:ed_benchmark_3x3}, using the combined post-selection and DEM of Sec.~\ref{sec:combination}. In the deconfined regime, the mitigated data closely track the exact-diagonalization reference. As $h_z/J_v$ increases toward the confined regime, the agreement on the gauge links degrades, which we attribute to the non-Clifford $h_z$ rotations: their angle grows with $h_z$, and they are absent from the Clifford circuit used to estimate the noise (Sec.~\ref{sec:implementation}).

\subsection{Local approximation of string-like projectors}

The projectors of Eq.~\eqref{eq:string_projector} are diagonal in the
measurement basis and hence directly accessible. However, combining local post-selection with marginal DEM requires a cluster containing the entire string and its neighborhood. Since this cluster grows with the string length and the reconstructed marginal distribution grows exponentially with the cluster size, the direct mitigation of string projectors becomes hardly scalable for longer strings spanning larger lattices.

While the full projectors remain accessible at the present string lengths, as shown in Fig.~\ref{fig:projectors}, longer strings and larger lattices will call for a more economical estimator. We therefore introduce an approximation that retains only two-link
correlations along the path. For a string $\gamma$ of $L$ links,
\begin{equation}
  \label{eq:string_approx}
  \Papprox_{\gamma}
  = \frac{\prod_{\langle i,j\rangle \in \gamma}\langle P_iP_j\rangle}
         {\prod_{i\in\gamma\setminus\partial\gamma}\langle P_i\rangle} ,
\end{equation}
where $P_i=(1-\sigma_i^z)/2$ is the occupation of link $i$, $\langle i,j\rangle$ denotes neighboring links along the path and
$\partial\gamma$ its two boundary links. The denominator compensates for the
overlap of adjacent two-link projectors, preserving the normalization, and
reduces the measurement requirement from an $L$-body projector to one- and
two-body expectation values. We mitigate each two-link projector on the local cluster of Fig.~\ref{fig:postselection}(C), which contains the $3$ Gauss stabilizers of the matter sites adjacent to the two links, and each single-link occupation on the cluster of Fig.~\ref{fig:postselection}(B).

\begin{figure*}[ht]
    \centering
    \includegraphics[width=\textwidth]{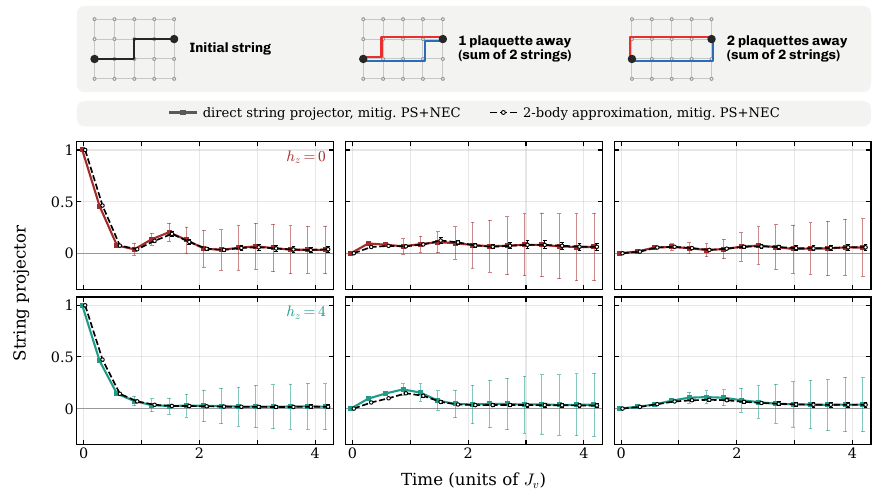}
    \caption{\textbf{Comparison of the directly mitigated string projectors}
    (Fig.~\ref{fig:projectors}) \textbf{with the two-body approximation}
    (Eq.~\eqref{eq:string_approx}), for the initial string (left) and summed over
    the two strings 1 plaquette (center) and 2 plaquettes (right) away, for
    $h_z=0$ and $h_z=4$. Error bars of the approximation are propagated
    assuming independent one- and two-body terms and do not include the
    systematic error of the approximation.}
    \label{fig:string_pair_approx}
\end{figure*}

Figure~\ref{fig:string_pair_approx} benchmarks this estimator against the
directly mitigated projectors for the five minimal-length strings of
Fig.~\ref{fig:projectors}. We find generally good agreement with the full projector in both the confined and deconfined regimes. Moreover, the statistical error bound of the distribution mitigation is significantly lower for the two-link projectors than for the full projector, due to their smaller marginal distributions. Since the cluster of each two-link projector does not grow with the string length, the approximation in principle allows a quantitative, mitigated measurement of string oscillations on larger lattices. 

\section{Error propagation, syndrome decoding and gauge protection}
\label{app:propagation}
\subsection{End-of-circuit syndromes}

Rather than discarding every corrupted shot, one may attempt to decode the
measured end-of-circuit syndromes and correct where possible, recovering a
fraction of the affected samples. Such an approach has been used to improve
signal recovery in a $2+1$D $\Ztwo$ simulation without a plaquette term on a
hexagonal-connectivity IBM
processor~\cite{cobosRealTimeDynamics2+1D2025}. We assess its potential here.

For measurement errors alone the situation is clean: assuming single-qubit
readout errors, the Gauss stabilizers unambiguously detect and correct bit flips
on both gauge and matter qubits, since a gauge-qubit flip flips two neighboring
Gauss stabilizers and a matter-qubit flip flips one.

Errors during the Trotter evolution are less obvious, since they propagate.
Tables~\ref{tab:prop_vertical} and~\ref{tab:prop_horizontal} record the
propagation and end-of-step syndromes of single-qubit $X$ and $Z$ errors on
gauge qubits as a function of the substep at which they occur, with substeps
labeled as in Fig.~\ref{fig:gauge_protection}. Each vertical gauge qubit
neighbors two Gauss stabilizers (upper and lower matter qubits) and two ancilla
stabilizers (left and right); each horizontal gauge qubit neighbors two Gauss
stabilizers (left and right) and two ancilla stabilizers (above and below). Throughout this analysis the evolution circuit is approximated by its underlying Clifford structure.

Two syndrome patterns decode unambiguously, independently of the substep:
\begin{itemize}
  \item simultaneous sign flips on two neighboring \emph{Gauss} stabilizers
        correspond to a single bit-flip error on the gauge qubit between them;
  \item simultaneous sign flips on two neighboring \emph{ancilla} stabilizers
        correspond to a single phase-flip error on the gauge qubit between them.
\end{itemize}
Since $Z$ errors are not directly visible in the computational basis, the first
is the relevant one for end-of-circuit correction. Single-qubit bit flips on ancilla and matter qubits do not spread, since both
appear only as CNOT targets; at the end of the circuit they produce single
Gauss-stabilizer flips that do not interfere with the unambiguous signature
above. Single-qubit phase flips on ancilla and matter qubits can, however,
propagate into errors of weight up to $6$, and can therefore in some cases
produce patterns overlapping with the gauge-qubit signature. One nonetheless
concludes that, provided there are no multiple errors within the same local
region, a pair of neighboring Gauss-stabilizer flips can be decoded with high
confidence.


\begin{table}[t]
\centering
\footnotesize
\setlength{\tabcolsep}{2.5pt}
\begin{tabular}{cccccccc c}
\toprule
\textbf{Err.} & \textbf{Step} & \textbf{R} & \textbf{L} &
\makecell{\textbf{U}\\\textbf{mid}} & \makecell{\textbf{D}\\\textbf{mid}} &
\makecell{\textbf{U}\\\textbf{end}} & \makecell{\textbf{D}\\\textbf{end}} &
\makecell{\textbf{Weight}} \\
\midrule
$I$ & --      & $Z$ & $Z$ & $Z$  & $Z$  & $ZZZZZ$  & $ZZZZZ$  & -- \\
\midrule
$X$ & 0--1    & $Z$ & $Z$ & $-Z$ & $-Z$ & $-ZZZZZ$ & $-ZZZZZ$ & 1 \\
$X$ & 2       & $Z$ & $Z$ & $Z$  & $-Z$ & $ZZZZZ$  & $-ZZZZZ$ & 2 \\
$X$ & 3--6    & $Z$ & $Z$ & $Z$  & $Z$  & $ZZZZZ$  & $ZZZZZ$  & 3 \\
$X$ & 7       & $Z$ & $Z$ & $Z$  & $Z$  & $ZZZZZ$  & $-ZZZZZ$ & 2 \\
$X$ & 8--10   & $Z$ & $Z$ & $Z$  & $Z$  & $-ZZZZZ$ & $-ZZZZZ$ & 1 \\
\midrule
$Z$ & 0       & $Z$  & $Z$  & $Z$ & $Z$ & $ZZZZZ$ & $ZZZZZ$ & 1 \\
$Z$ & 1--3    & $-Z$ & $Z$  & $Z$ & $Z$ & $ZZZZZ$ & $ZZZZZ$ & 2 \\
$Z$ & 4--5    & $-Z$ & $-Z$ & $Z$ & $Z$ & $ZZZZZ$ & $ZZZZZ$ & 3 \\
$Z$ & 6--8    & $-Z$ & $Z$  & $Z$ & $Z$ & $ZZZZZ$ & $ZZZZZ$ & 2 \\
$Z$ & 9--10   & $Z$  & $Z$  & $Z$ & $Z$ & $ZZZZZ$ & $ZZZZZ$ & 1 \\
\bottomrule
\end{tabular}
\caption{Propagation of single-qubit Pauli-$X$ and Pauli-$Z$ errors on a
\emph{vertical} gauge qubit, by Trotter substep (numbering as in
Fig.~\ref{fig:gauge_protection}). Errors after the last substep may be read as
measurement errors. $R$, $L$ are the right and left ancilla stabilizers, $U$,
$D$ the upper and lower Gauss stabilizers; ``mid'' and ``end'' indicate the
midpoint and end of the Trotter step. The last column gives the end-of-circuit
error weight, i.e.\ the number of qubits the error has spread to.}
\label{tab:prop_vertical}
\end{table}

\begin{table}[t]
\centering
\footnotesize
\setlength{\tabcolsep}{2.5pt}
\begin{tabular}{cccccccc c}
\toprule
\textbf{Err.} & \textbf{Step} & \textbf{U} & \textbf{D} &
\makecell{\textbf{R}\\\textbf{mid}} & \makecell{\textbf{L}\\\textbf{mid}} &
\makecell{\textbf{R}\\\textbf{end}} & \makecell{\textbf{L}\\\textbf{end}} &
\makecell{\textbf{Weight}} \\
\midrule
$I$ & --      & $Z$ & $Z$ & $Z$  & $Z$  & $ZZZZZ$  & $ZZZZZ$  & -- \\
\midrule
$X$ & 0       & $Z$ & $Z$ & $-Z$ & $-Z$ & $-ZZZZZ$ & $-ZZZZZ$ & 1 \\
$X$ & 1--3    & $Z$ & $Z$ & $Z$  & $-Z$ & $ZZZZZ$  & $-ZZZZZ$ & 2 \\
$X$ & 4--5    & $Z$ & $Z$ & $Z$  & $Z$  & $ZZZZZ$  & $ZZZZZ$  & 3 \\
$X$ & 6--8    & $Z$ & $Z$ & $Z$  & $Z$  & $ZZZZZ$  & $-ZZZZZ$ & 2 \\
$X$ & 9--10   & $Z$ & $Z$ & $Z$  & $Z$  & $-ZZZZZ$ & $-ZZZZZ$ & 1 \\
\midrule
$Z$ & 0--1    & $Z$  & $Z$  & $Z$ & $Z$ & $ZZZZZ$ & $ZZZZZ$ & 1 \\
$Z$ & 2       & $Z$  & $-Z$ & $Z$ & $Z$ & $ZZZZZ$ & $ZZZZZ$ & 2 \\
$Z$ & 3--6    & $-Z$ & $-Z$ & $Z$ & $Z$ & $ZZZZZ$ & $ZZZZZ$ & 3 \\
$Z$ & 7       & $Z$  & $-Z$ & $Z$ & $Z$ & $ZZZZZ$ & $ZZZZZ$ & 2 \\
$Z$ & 8--10   & $Z$  & $Z$  & $Z$ & $Z$ & $ZZZZZ$ & $ZZZZZ$ & 1 \\
\bottomrule
\end{tabular}
\caption{As Table~\ref{tab:prop_vertical}, for a \emph{horizontal} gauge qubit.
Here $U$, $D$ are the upper and lower ancilla stabilizers and $R$, $L$ the right
and left Gauss stabilizers.}
\label{tab:prop_horizontal}
\end{table}

\subsection{How reliable is end-of-circuit decoding?}

To test this in a more realistic setting we simulate the Clifford circuit
corresponding to $(h_z,h_x,J_v,J_v)=(0,0,0,0)$ with a Pauli noise channel
inserting single-qubit Pauli errors with probability $p$ after each CNOT, and
measure how often the observed syndrome patterns are indeed caused by the
single-qubit process one would infer. We consider two patterns: a pair of
neighboring Gauss-stabilizer flips, attributed to a bit flip on the gauge qubit
between them; and an isolated single Gauss-stabilizer flip, attributed to a bit
flip on the corresponding matter qubit. Only the first is expected to be
near-unambiguous, since isolated flips can also arise from errors on neighboring
gauge qubits that propagate to the same final pattern.

\begin{figure}[t]
  \centering
  \includegraphics[width=0.6\linewidth]{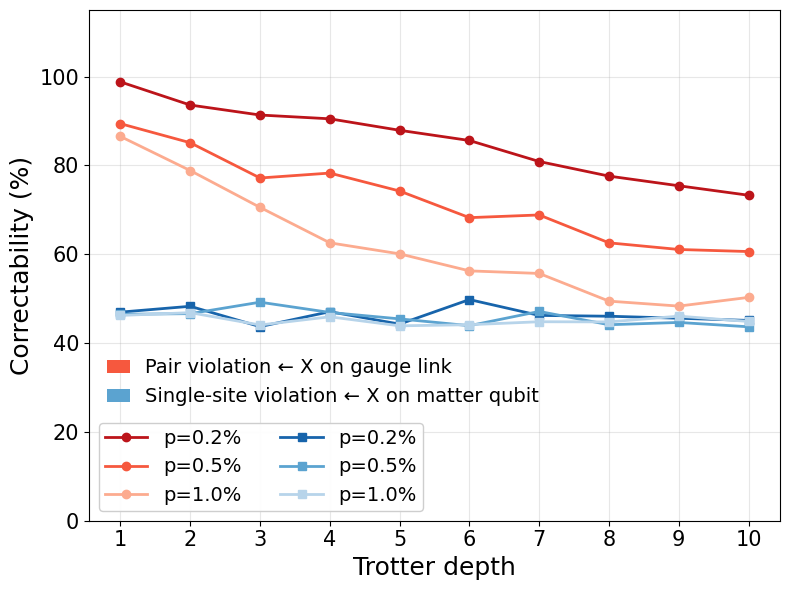}
  \caption{Probability that an observed end-of-circuit syndrome pattern is
  generated by the single-qubit process described in the legend, versus Trotter
  depth, for the Clifford circuit of a $5\times4$ lattice, at three single-qubit
  Pauli error rates with errors inserted after each CNOT.}
  \label{fig:correctability}
\end{figure}

Figure~\ref{fig:correctability} shows the results for a $5\times4$ lattice. At
low error rates, pairwise flips are predominantly caused by single bit flips on
the intervening gauge qubit even at large Trotter depth. As the error rate
grows, the scheme degrades toward the regime where applying the corresponding
correction introduces about as many errors as it removes. For isolated
single-stabilizer flips the probability of the inferred origin sits at
$\approx50\%$ at all depths, so correcting on that basis is as likely to
introduce an error as to fix one and is not a useful strategy.

We conclude that end-of-circuit syndrome decoding reliably corrects single-qubit
measurement errors, and---for the pairwise Gauss pattern and sufficiently low
error rates---can also help for errors occurring during the evolution, extending
the accessible times relative to post-selection alone.

\subsection{Mid-circuit detection and gauge protection}

Error detection can also be performed at intermediate stages, catching errors
before they propagate further. Mid-circuit measurements are not yet reliably
supported on current hardware, given their duration and associated errors, but
our circuit accommodates them at no qubit or gate overhead: the ancilla
stabilizers are encoded on single qubits at the end of each Trotter step,
Fig.~\ref{fig:model}(C), and the Gauss stabilizers on single matter qubits at
its midpoint, Fig.~\ref{fig:model}(D). Both are therefore accessible by
single-qubit measurements at natural points of the circuit.

Beyond detecting incoherent errors, mid-circuit measurement suppresses the
coherent accumulation of unitary gate errors, which may be viewed as unwanted
contributions to an effective Hamiltonian,
\begin{equation}
  \Heff = H - \beta H_{\mathrm{err}} ,
  \qquad [H_{\mathrm{err}},G_i]\neq 0 ,
\end{equation}
with $\beta$ the error strength. Projecting the wavefunction back into the
stabilized subspace acts as dynamical gauge protection; above a sufficiently high
measurement rate this suppresses gauge-breaking coherent errors, with the
transition occurring through a sharp quantum Zeno phase
transition~\cite{wautersSymmetryprotectionZenoPhase2025}.

\paragraph*{Energy penalty.} Adding a term penalizing Gauss-law violations,
\begin{equation}
  H_{\mathrm{prot}} = H - \beta H_{\mathrm{err}} - \alpha \sum_i G_i ,
\end{equation}
shifts gauge-violating states by $\propto\alpha$ and suppresses transitions out
of the physical subspace; for $\alpha$ sufficiently large compared with $\beta$
the dynamics is effectively
constrained~\cite{halimehGaugeSymmetryProtectionUsing2021}.

\paragraph*{Pseudo-random gauge transformations.} Alternatively one inserts
gauge transformations into the Trotter product of
Eq.~\eqref{eq:trotter}~\cite{cobosRealTimeDynamics2+1D2025},
\begin{equation}
  \tilde U(t) = \prod_{k=1}^{N}
    U_A(dt/2)\,U_B(dt)\,U_G(\{\phi_{k,n}\})\,U_A(dt/2) ,
\end{equation}
with $U_G(\{\phi_{k,n}\})=\prod_n e^{-i\phi_{k,n}G_n}$ and random angles
$\phi_{k,n}$ chosen such that $\sum_k\phi_{k,n}=0$, so no systematic coherent
phase bias is introduced. Suppose that after a Trotter step coherent gate
imperfections have generated a small gauge-violating amplitude,
$|\psi\rangle=|\psi_{\mathrm{phys}}\rangle+\epsilon|\psi_{\mathrm{unphys}}\rangle$,
with $G_n|\psi_{\mathrm{phys}}\rangle=+|\psi_{\mathrm{phys}}\rangle$ and
$G_n|\psi_{\mathrm{unphys}}\rangle=g_n|\psi_{\mathrm{unphys}}\rangle$,
$g_n=\pm1$, with $g_n\neq1$ for at least one $n$. Then, up to a global phase,
\begin{equation}
  |\tilde\psi\rangle = |\psi_{\mathrm{phys}}\rangle
    + \epsilon e^{-i\Delta\Phi}|\psi_{\mathrm{unphys}}\rangle ,
  \quad \Delta\Phi = \sum_n \phi_{k,n}(g_n-1) .
\end{equation}
The off-diagonal terms of the corresponding density matrix carry
$e^{\pm i\Delta\Phi}$; since the angles vary between circuit realizations,
$\langle e^{\pm i\Delta\Phi}\rangle\approx0$ and the ensemble average becomes
\begin{equation}
  \langle\rho\rangle \approx
    |\psi_{\mathrm{phys}}\rangle\langle\psi_{\mathrm{phys}}|
    + \epsilon^2|\psi_{\mathrm{unphys}}\rangle\langle\psi_{\mathrm{unphys}}| .
\end{equation}
Applying independent random transformations at every step continuously
randomizes the phases of newly generated gauge-violating amplitudes, so leakage
generated at different times no longer interferes constructively and the
coherent accumulation is suppressed. Related schemes based on dynamical
decoupling have also been
proposed~\cite{kasperNonAbelianGaugeInvariance2023}.

\begin{figure*}[t]
  \centering
  \includegraphics[width=0.95\linewidth]{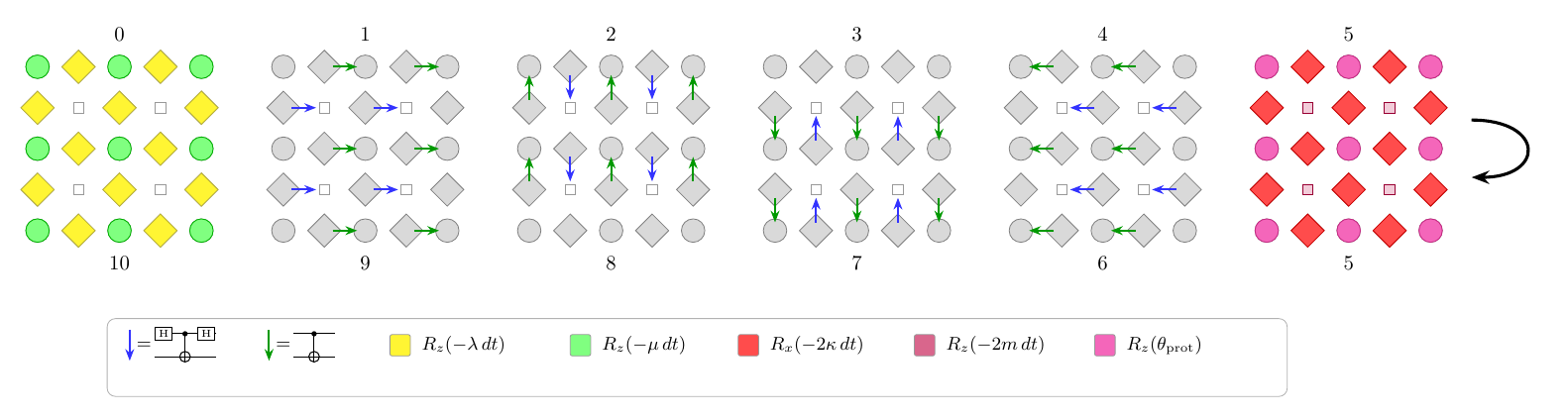}
  \caption{Implementation of gauge protection through single-qubit
  $R_z(\theta_{\mathrm{prot}})$ rotations on the matter qubits at the midpoint
  of each Trotter step, where the Gauss-law generators are encoded into
  single-qubit $Z$ operators. Here $\theta_{\mathrm{prot}}=-2\alpha\,dt$ for the
  energy-penalty scheme and $\theta_{\mathrm{prot}}=-2\phi_{k,n}$ for the
  pseudo-random gauge transformation scheme. The circuit substeps are labeled
  $0$--$10$, matching Tables~\ref{tab:prop_vertical}
  and~\ref{tab:prop_horizontal}.}
  \label{fig:gauge_protection}
\end{figure*}

\paragraph*{Implementation.} Both schemes are free in our construction. Since
the generators $G_i$ are encoded into single matter qubits at the midpoint of
each Trotter step, both reduce to single-qubit $R_z(\theta_{\mathrm{prot}})$
rotations on the matter qubits at that point, Fig.~\ref{fig:gauge_protection},
with $\theta_{\mathrm{prot}}=-2\alpha\,dt$ for the energy penalty and
$\theta_{\mathrm{prot}}=-2\phi_{k,n}$ for the random gauge transformation.

\section{Classical simulability and entanglement}
\label{app:simulability}

Quantum Monte Carlo is excluded for real-time dynamics due to the sign
problem~\cite{QuarksStringsLattice}, and the state-of-the-art classical approach is based on tensor networks~\cite{Verstraete2004RenormalizationAlgorithmsQuantumManya,orusTensorNetworksComplex2019}, whose
efficiency is tied to the entanglement of the evolving state. The optimal TN representation varies across
dynamical regimes and excitation geometries, so that geometry, bond dimension or
contraction strategy must be adapted case by case. In two dimensions this is
sharper still~\cite{magnificoTensorNetworksLattice2025}: matrix product
states~\cite{ostlundThermodynamicLimitDensity1995} require mapping the lattice
onto an effective chain, with efficiency depending strongly on the correlation
structure~\cite{scardicchioOptimalLayoutTwoDimensional2026}, while projected
entangled pair states preserve the geometry
naturally~\cite{kelmanProjectedEntangledPair2024} but remain expensive to
contract and evolve~\cite{schuchComputationalComplexityPEPS2007}. A state-of-the-art reference for the dynamics would use Belief-Propagation for contracting PEPS tensor networks expectation values, although we leave this to further work as our main current limitation is the low bond-dimension of the quantum state itself.

\end{document}